# Probing the Interplay between Quantum Charge Fluctuations and Magnetic Ordering in LuFe$_2$O$_4$

J. Lee[1*], S. A. Trugman[1,2], C. D. Batista[2], C. L. Zhang[3], D. Talbayev[4], X. S. Xu[5], S.–W. Cheong[3], D. A. Yarotski[1], A. J. Taylor[1], and R. P. Prasankumar[1#]

[1]*Center for Integrated Nanotechnologies, Los Alamos National Laboratory, Los Alamos, NM 87545*
[2]*Theoretical Division, Los Alamos National Laboratory, Los Alamos, NM 87545*
[3]*Rutgers Center for Emergent Materials and Department of Physics and Astronomy, Rutgers University, Piscataway, NJ 08854*
[4]*Department of Physics and Engineering Physics, Tulane University, New Orleans, LA 70118*
[5]*Oak Ridge National Laboratory, Oak Ridge, TN 37831*
[*]*Email: kjindda@naver.com*
[#]*Email: rpprasan@lanl.gov*

**Ferroelectric and ferromagnetic materials possess spontaneous electric and magnetic order, respectively, which can be switched by applied electric and magnetic fields. Multiferroics combine these properties in a single material, providing an avenue for controlling electric polarization with a magnetic field and magnetism with an electric field. These materials have been intensively studied in recent years, both for their fundamental scientific interest as well as their potential applications in a broad range of magnetoelectric devices [1, 2, 3, 4]. However, the microscopic origins of magnetism and ferroelectricity are quite different, and the mechanisms producing strong coupling between them are not always well understood. Hence, gaining a deeper understanding of magnetoelectric coupling in these materials is the key to their rational design. Here, we use ultrafast optical spectroscopy to show that the influence of magnetic ordering on quantum charge fluctuations via the double-exchange mechanism can govern the interplay between electric polarization and magnetism in the charge-ordered multiferroic LuFe$_2$O$_4$.**

Recently, the iron-based multiferroic LuFe$_2$O$_4$ has attracted much attention because it exhibits magnetoelectric coupling close to room temperature [5~21]. The unique layered structure of LuFe$_2$O$_4$ consists of double layers of Fe ions connected in a triangular lattice in the *ab*-plane (Figure 1(a)) [14]. The average valence of Fe ions is Fe$^{2.5+}$, with Fe$^{2+}$ and Fe$^{3+}$ ions occupying equivalent sites in different layers with equal densities. The corresponding spin values are $S=2$ (Fe$^{2+}$) and $S=5/2$ (Fe$^{3+}$), with the spin structure shown in Figure 1(b) [15,16]. A simple description based on nearest-neighbour interactions between Fe ions leads to the characterization of this material as a spin and charge frustrated system [5, 17]. Bulk ferroelectricity was observed below the charge ordering temperature, $T_{CO}\sim$ 320 K, resulting in a spontaneous electric polarization that further increased upon the appearance of ferrimagnetic spin order below the Neel temperature, $T_N\sim$240 K [5]. Ferroelectricity in each bilayer is thus induced by electronic charge ordering, although the stacking of adjacent bilayers (i.e., in an antiferroelectric or ferroelectric arrangement) is still controversial [11,12,15,18]. Regardless, in each bilayer, the electric polarization **P** is coupled to the magnetic degrees of freedom in LuFe$_2$O$_4$, but a comprehensive understanding of the mechanism underlying this magnetoelectric coupling has eluded researchers to date. Knowledge of this mechanism could potentially allow researchers to optimize both the strength of this coupling and its operating temperature to address the general goal of developing multiferroic materials with strong magnetoelectric coupling at room temperature.

Theoretical studies have linked the magnetoelectric coupling in LuFe$_2$O$_4$ to both thermal [19] and quantum [20, 21] charge fluctuations. In general, magnetic ordering can modify the effective hopping amplitude between two ions via the well-known double-exchange mechanism [20~23], in which hopping is governed by the angle between the two core spins, as shown in Figure 1(b). This change in the hopping amplitude will necessarily affect the quantum fluctuations of any charge ordered state of electronic origin. The corresponding change in charge ordering will necessarily modify **P**. In other words, if electronic charge ordering leads to a net electric polarization, the value of **P** should be modified by the presence of magnetic ordering (Figure 1(b)). Moreover, if the magnetic ordering *reduces* (on average) the effective hopping amplitude, the corresponding suppression of quantum charge fluctuations leads to an increase of **P** below $T_N$. Since the same mechanism should affect optically induced charge fluctuations, we use femtosecond optical pump-probe spectroscopy, which has been extensively used to shed light on the properties of correlated electron materials [24~31, to directly photoexcite and probe the $Fe^{2+} \rightarrow Fe^{3+}$ charge transfer channel in $LuFe_2O_4$. Then, by varying the temperature $T$ above and below $T_N$, we can shed light on the role of these fluctuations in governing the coupling between spin and charge order in a single $LuFe_2O_4$ bilayer, regardless of whether the bilayers are stacked ferroelectrically or antiferroelectrically. We find that the interlayer hopping matrix element describing these fluctuations depends strongly on their local core spin alignment via the double-exchange mechanism, making charge delocalization (in real space, as shown in Figure 1(b)) and hence the electric polarization extremely sensitive to the spin structure evolution over a broad temperature range. Therefore, although magnetoelectric coupling in various multiferroic materials has been studied using many different techniques [32], to the best of our knowledge, this is the first experimental evidence of magnetoelectric coupling mediated by the double-exchange mechanism in an insulator.

We begin by developing a model for electronic hopping between two atomic sites, governed by the double exchange mechanism, which shows that the transition rate (i.e., charge transfer rate) between the ground and the excited state induced by an external driving electromagnetic field is proportional to the effective hopping matrix, $t_{ij}^2$, according to the Fermi Golden rule [33], as described in the Methods section. Furthermore, the amount of delocalized charge $\delta q$ is proportional to $t_{ij}^2$, $\delta q = (\frac{t_{ij}}{\Delta})^2$, where $\Delta$ is the energy difference between the ground and excited states. This simple observation establishes our ability to probe quantum charge fluctuations between two atomic sites using optical spectroscopy (see Methods for more detail). In $LuFe_2O_4$, these fluctuations are due to charge transfer between $Fe^{2+}$ and $Fe^{3+}$ ions (as revealed by optical spectroscopy [34] and band structure calculations [19]). We can consider four different charge transfer channels in the bilayer crystal structure of $LuFe_2O_4$,: interlayer charge transfer from the $Fe^{2+}$ rich bottom layer to the $Fe^{3+}$ rich top layer ($E_\uparrow$) or from the top to the bottom layer ($E_\downarrow$), and intralayer charge transfer within the top layer ($E_{t\rightarrow}$) and within the bottom layer ($E_{b\rightarrow}$), as shown in Figure 1(a). We can gain insight on the relative energies of these different charge transfer channels by considering the Coulomb energy between Fe ions in the Hamiltonian,

$$H_V = \sum_{(ij)} \frac{Q_i^z Q_j^z}{4\pi\varepsilon_r \varepsilon_0 r_{ij}}, \tag{1}$$

where the pseudospin operators $Q_i^z$ and $Q_j^z$ are 1/2 or -1/2 for $Fe^{3+}$ or $Fe^{2+}$, respectively, and $\varepsilon_0, \varepsilon_r$ and $r_{ij}$ are the permittivity of free space, the relative permittivity and the distance between sites $i$ and $j$, respectively. Considering only the largest three interaction terms, we find that $E_\uparrow$ has the lowest excitation energy, $E_{t\rightarrow}$ and $E_{b\rightarrow}$ have intermediate excitation energies, and $E_\downarrow$ has the highest excitation energy. The interlayer transitions can be distinguished by noting that the bottom layer is rich in $Fe^{2+}$ while the top layer is rich in $Fe^{3+}$. It is clear that if the top layer has a positive charge density σ>0 per unit area, the bottom layer must have the opposite charge density, –σ, to ensure charge neutrality. $E_↓$ increases with σ while $E_↑$ decreases, so it is reasonable to assume that $E_↓ >> E_↑$ in LuFe$_2$O$_4$. We can also distinguish the intralayer transitions by noting that the configuration of in-plane oxygen ions around Fe ions in both layers leads to a higher in-plane charge transfer excitation energy for $E_{b→}$ than that of $E_{t→}$. The optical conductivity measurements described in ref. [34] show two distinct charge transfer excitation channels at ~1.1 eV and ~1.5 eV, which should thus correspond to $E_↑$ and $E_{t→}$, respectively. To further confirm this, we performed angle-dependent reflectivity measurements (not shown), which agreed well with the data in ref. [34] and allowed us to verify that $E_↑$ ~1.1 eV and $E_{t→}$ ~1.5 eV by tracking the strength of these absorption peaks as a function of angle and polarization. It is worth noting that both in our measurements and in the data of ref. [34], no spectral signatures corresponding to $E_{b→}$ and $E_↓$ were observed. This is likely because there are many different possible transitions that overlap at higher energies, which obscure the peaks corresponding to $E_{b→}$ and $E_↓$. Therefore, we used photon energies of 1.1 ($E_↑$) and 1.5 ($E_{t→}$) eV in our experiments to examine inter- and intralayer quantum charge fluctuations in LuFe$_2$O$_4$. This is actually advantageous, since the direction of **P** in LuFe$_2$O$_4$ is nearly parallel to the *c*-axis with a small angle (Fig. 1), directly linking it to the 1.1 eV interlayer charge fluctuations [20, 21].

We propose that magnetic order and charge fluctuations in LuFe$_2$O$_4$ are linked through the double exchange mechanism[20, 21], which leads to an effective hopping matrix element $t_{ij}$ (see Methods) between the ions *i* and *j* that is determined by the angle $θ_{ij}$ between the spins **S**$_i$ and **S**$_j$: $t_{ij} = t\cos(θ_{ij}/2)$ [23]. Thermal fluctuations prevent any preferred spin orientation in the paramagnetic phase ($T>T_N$). Within an Ising spin model, in the paramagnetic phase the hopping matrix element $t_{ij}$ will 0 (if the $Fe^{2+}$ and $Fe^{3+}$ core spins are antiparallel) or $t$(if they are parallel) and thus its average value will be $t$/2 for this transition, and similarly, the average value of $t_{ij}^2$ is $t_{ij}^2$/2. However, in the magnetically ordered state ($T<T_N$), $t_{ij}=t$ if the nearest neighbour spins are aligned ferromagnetically (FM), while $t_{ij}=0$ if they are antiferromagnetically (AFM) aligned (Fig. 1(b) and Fig. 2). This can then be applied to the four charge transfer channels discussed above (considering electron hopping from a given $Fe^{2+}$ ion to its nearest neighbour $Fe^{3+}$ ions) to find the total steady-state absorption (proportional to $t_{ij}^2$) for each transition above and below $T_N$. We find that the development of magnetic order does not affect the total $t_{ij}^2$ for $\boldsymbol{E}_\uparrow$ and $\boldsymbol{E}_{t\rightarrow}$. This can straightforwardly be seen by considering the $\boldsymbol{E}_\uparrow$ transition above $T_N$ (Figure 1(a)), where the spins do not have a preferred direction and below $T_N$ . . Because there are two possible charge transfer channels for $\boldsymbol{E}_\uparrow$(Figure 1(a)) , we obtain a total transition amplitude of $t^2/2\times2=t^2$ above $T_N$ with average value of $t^2$ /2 as described in above. Below $T_N$, there is only one possible transition with an amplitude of $t^2$ (Figure 1(b)), so the transition amplitudes above and below $T_N$ are the same. In the same manner, if we consider the $E_{t\rightarrow}$ transition above $T_N$ (Figure 1(a)), there are six charge transfer channels, giving a transition amplitude of $t^2/2\times6=3t^2$. Below $T_N$, there are only three possible transitions, each with amplitude $t_{ij}^2$, so the transition amplitudes above and below $T_N$ are the same, as for the interlayer transition.

This agrees with the data in ref. [34], which shows that the absorption for the $\boldsymbol{E}_\uparrow$ and $\boldsymbol{E}_{t\rightarrow}$ transitions does not significantly change with temperature. In contrast, the total $t_{ij}^2$ is reduced through the double exchange mechanism for $\boldsymbol{E}_{b\rightarrow}$ and $\boldsymbol{E}_\downarrow$ below $T_N$; therefore, as described in the introduction, the corresponding suppression of quantum charge fluctuations should increase **P**. A simple test of the influence of the double exchange mechanism on magnetoelectric coupling in LuFe$_2$O$_4$ would thus be to track the expected changes in the steady-state optical absorption at $E_{b\rightarrow}$ and $E_{\downarrow}$ as the temperature $T$ is varied across $T_N$. However, as described above, this is not possible since the energies of these transitions are unknown and couldn't find any specific peaks related with these transitions in the absorption spectrum as shown in ref. [34].

One can circumvent this limitation by performing a *non-equilibrium* experiment; i.e., photoexciting the known $E_{\uparrow}$ and $E_{t\rightarrow}$ transitions and examining the resulting changes in the optical absorption at the $E_{\uparrow}$ transition (which is proportional to **P**). Photoexcitation changes the charge configuration, which in turn changes both the total $t_{ij}^2$ and the $E_{\uparrow}$ charge transfer energy through equation (1), leading to a transient change in the reflectivity that can be measured in our experiments (see Methods for more detail). The development of magnetic order below $T_N$ can then further modify the absorption at $E_{\uparrow}$ through the double exchange mechanism in the same manner as in the steady state, potentially causing an additional change in the photoinduced reflectivity. In short, our ultrafast optical experiments (described in more detail in Methods) allow us to examine the effect of photoexciting either intralayer or interlayer charge transfer transitions (in effect externally driving charge fluctuations) on the interlayer charge transfer energy as a function of temperature. Tuning the temperature above and below $T_N$ then allows us to determine the effect of magnetic ordering on charge fluctuations. In this way, we can test if these fluctuations are indeed responsible for the magnetoelectric coupling measured in LuFe$_2$O$_4$.

Figure 3(a) shows the temporal profile of the normalized photoinduced reflectivity change, $\Delta R/R(t)$, in LuFe$_2$O$_4$ at several different temperatures for a degenerate 1.1 eV pump-probe measurement. Immediately after photoexcitation, $\Delta R/R$ decreases to its minimum value within <0.5 picoseconds (ps) (Figure 3(b)) and returns to equilibrium while exhibiting coherent acoustic phonon oscillations with a period of ~40 ps (Fig. 3(a)). These coherent phonons are generated by the dynamic stress on the sample induced by absorption of the pump pulses and are commonly observed in ultrafast optical experiments on correlated electron materials [30], as well as non-correlated materials described by a simple band structure [35]. In this paper, we will focus on the variation of the maximum amplitude of the transient reflectivity signal ($\Delta R/R_{max}$) with temperature, which gives insight into the influence of intra- and interlayer charge fluctuations on magnetoelectric coupling in $LuFe_2O_4$.

Figure 3(c) depicts $\Delta R/R_{max}$ (1.1 eV) after 1.1 eV photoexcitation as a function of temperature near $T_N$. As described above, photoexcitation changes the charge distribution in $LuFe_2O_4$, altering the energy required to transfer a charge from the bottom to the top layer ($E_\uparrow$) (and thus the absorption/reflectivity probed at 1.1 eV) (Figure 2 (a)). We calculated $\Delta E_\uparrow$, the difference in energy before and after photoexcitation, for $T<T_N$ and $T>T_N$, using equation (1) (and using $\varepsilon_r=2$ from [34]) (see Methods). This calculation reveals that there is no change in the $\Delta R/R_{max}$ signal as $T$ is varied across $T_N$, which is consistent with our experimental observation (Figure 3(c)); fundamentally, although the specific allowed charge transfer transitions changes after photoexcitation and give $\Delta E_\uparrow$ or $dR/R$ signal as shown in figure 3(a) and (b), absorption change due to double exchange mechanism(Fig. 2(a)) across $T_N$ is very small and thus undetectable in our experiment. The negative sign of the signal is also expected since photoexcitation reduces the absorption at 1.1 eV.

Figure 4(a) shows the transient reflectivity change for the 1.1 eV interlayer transition after photoexciting the intralayer ($E_{t\rightarrow}$) charge transfer channel at 1.5 eV. The time-dependent dynamics are similar to those observed after photoexciting $E_\uparrow$, but the variation of $\Delta R/R_{max}$ with temperature is quite different (Figure 4(b)); in particular, a significant increase in the amplitude is clearly observed as the temperature rises above $T_N$. As described above, the steady state absorption for the 1.1eV ($E_\uparrow$) interlayer transition does not change across $T_N$; however, the probe absorption at this transition can change after 1.5 eV photoexcitation as $T$ is varied across $T_N$. To understand this, we calculated the effect of the photoinduced intralayer charge transfer at 1.5 eV on $E_\uparrow$ in the same manner as described above for $E_\uparrow$ (see Methods). Our calculation shows that the ratio of the maximum photoinduced change in reflectivity between $T<T_N$ and $T>T_N$ is ~0.9, which agrees very well with our experimental results (~0.87) (Figure 4). From this experimental observation, we deduce that the ferrimagnetic order influences the fluctuations of the charge ordered state that is responsible for the electric polarization in LuFe$_2$O$_4$ through the double exchange interaction. This result indicates that the interplay between charge fluctuations and magnetic ordering can result in strong magnetoelectric coupling at the Neel temperature. Finally, it is worth noting that this mechanism will generate an electric polarization in each bilayer, regardless of whether the ground state consists of layers stacked with ferroelectric or antiferroelectric order.

In summary, we used femtosecond optical pump-probe spectroscopy to investigate the role of the double exchange mechanism in the magnetoelectric coupling observed in LuFe$_2$O$_4$. Our experiments revealed that optically induced charge fluctuations are affected by magnetic order in a manner that is consistent with this mechanism. Importantly, this result opens an alternative route for finding strong magnetoelectric effects: charge ordering in transition metal oxides can naturally lead to electric polarization that is coupled to the magnetic degree of freedom via the double-exchange interaction.

**Acknowledgments:**

This work was performed at the Center for Integrated Nanotechnologies, a US Department of Energy, Office of Basic Energy Sciences (BES) user facility and under the auspices of the Department of Energy, Office of Basic Energy Sciences, Division of Material Sciences. Los Alamos National Laboratory, an affirmative action equal opportunity employer, is operated by Los Alamos National Security, LLC, for the National Nuclear Security administration of the U.S. Department of Energy under contract no. DE-AC52-06NA25396.

35. Liu R. *et. al.* Femtosecond pump-probe spectroscopy of propagating coherent acoustic phonons in $In_xGa_{1-x}N/GaN$ heterostructures. *Phys. Rev. B* **72**, 195335 (2005).

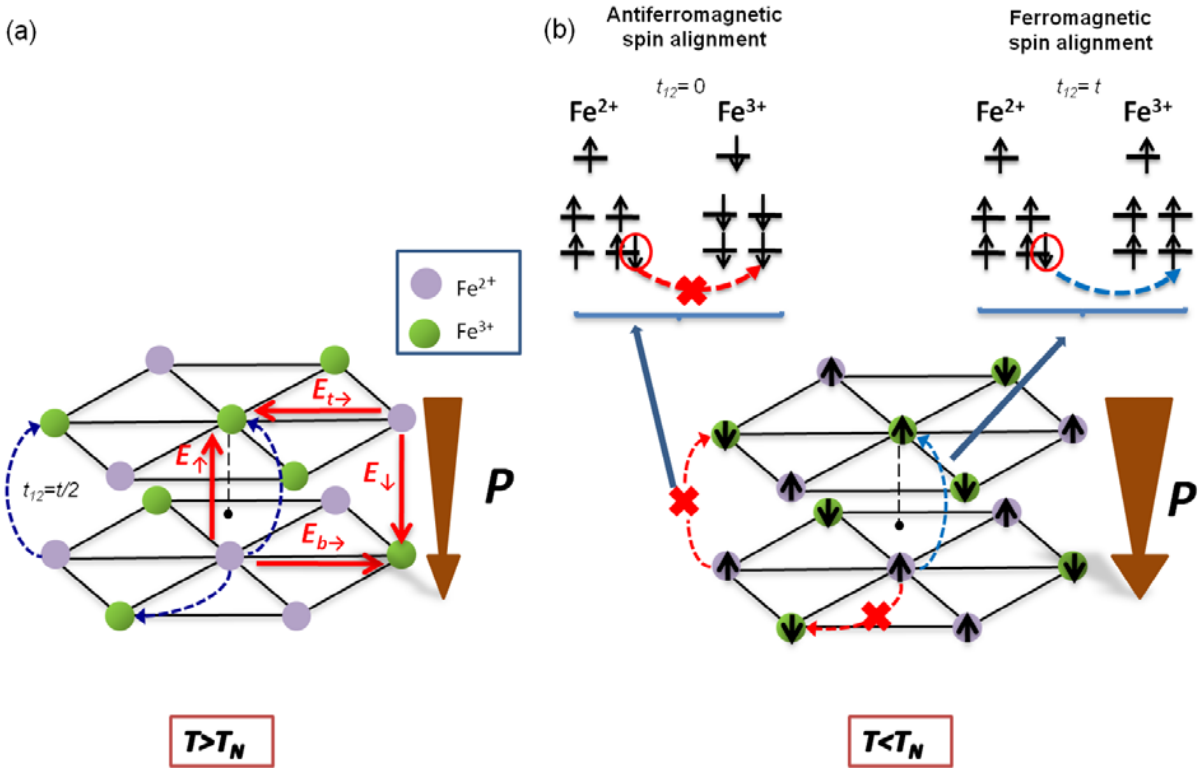

**Figure 1. Charge and spin ordering in $LuFe_2O_4$ above and below $T_N$.** (a) For $T>T_N$, charge ordering results in a finite polarization *P*. The top layer is displaced from the bottom layer by an angle shown by the black straight dashed line, which shows an iron atom in the upper plane directly above the center of a triangle in the lower plane. Quantum fluctuations between $Fe^{2+}$ and $Fe^{3+}$ ions (depicted by blue dashed lines) can reduce *P* by delocalizing charges, with an effective matrix element for hopping between two sites given by $t_{12}=t/2$. Red arrows show the possible charge transfer routes between $Fe^{2+}$ and $Fe^{3+}$, as defined in the text. (b) For $T<T_N$,

ferrimagnetic spin ordering increases *P* by decreasing the average hopping through the double exchange mechanism; $t_{12}=t$ when spins at both sites are ferromagnetically aligned (blue dashed line), and $t_{12}=0$ when both spins are antiferromagnetically aligned (red dashed lines).

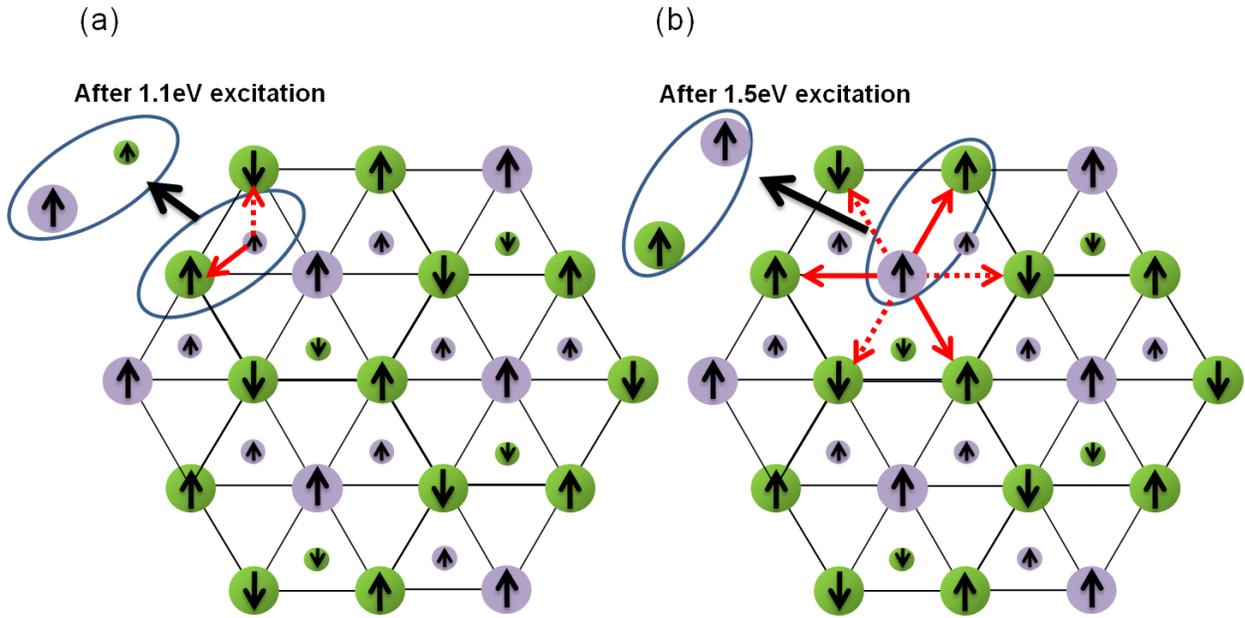

**Figure 2. Possible interlayer and intralayer charge transfer transitions for $T<T_N$.** Both figures show the spin structure of $LuFe_2O_4$ for $T<T_N$. Big and small solid circles correspond to atoms in the top and bottom layers, respectively. The black arrows show the local spins along the *c* axis at 220 K, from refs. [15] and [16]. Red solid and broken arrows represent allowed and forbidden interlayer (a) and intralayer (b) charge transfer channels, respectively, when considering the double exchange mechanism. After 1.1 eV excitation, one electron from an $Fe^{2+}$ ion in the bottom layer moves to an $Fe^{3+}$ ion in the top layer, while after 1.5 eV excitation, one

electron from an $Fe^{2+}$ site in the top layer moves to an $Fe^{3+}$ site in the same layer.

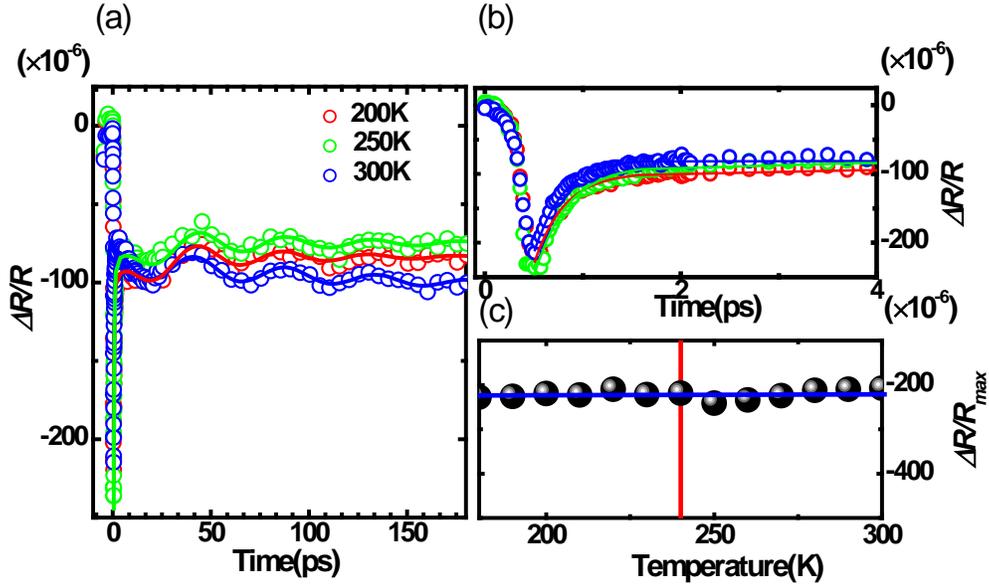

**Figure 3. Temperature-dependent transient reflectivity change after 1.1 eV photoexcitation.** (a) Temperature-dependent transient reflectivity change after photoexciting and probing the $Fe^{2+}$→$Fe^{3+}$ interlayer charge transfer excitation at 1.1 eV. The open circles are the experimentally obtained data points and the solid lines are the results of fitting the data with exponential and oscillating terms. (b) The data from (a) at early times. (c) The amplitude of the negative peak as a function of temperature. The red vertical line shows the magnetic transition temperature, $T_N$~240 K, and the blue horizontal line is parallel to the $x$ axis, which indicates that there is no significant variation in $\Delta R/R_{max}$ (1.1 eV) across $T_N$. We note that the high signal-to-noise ratio of our experiment resulted in error bars that are comparable to the size of the data points.

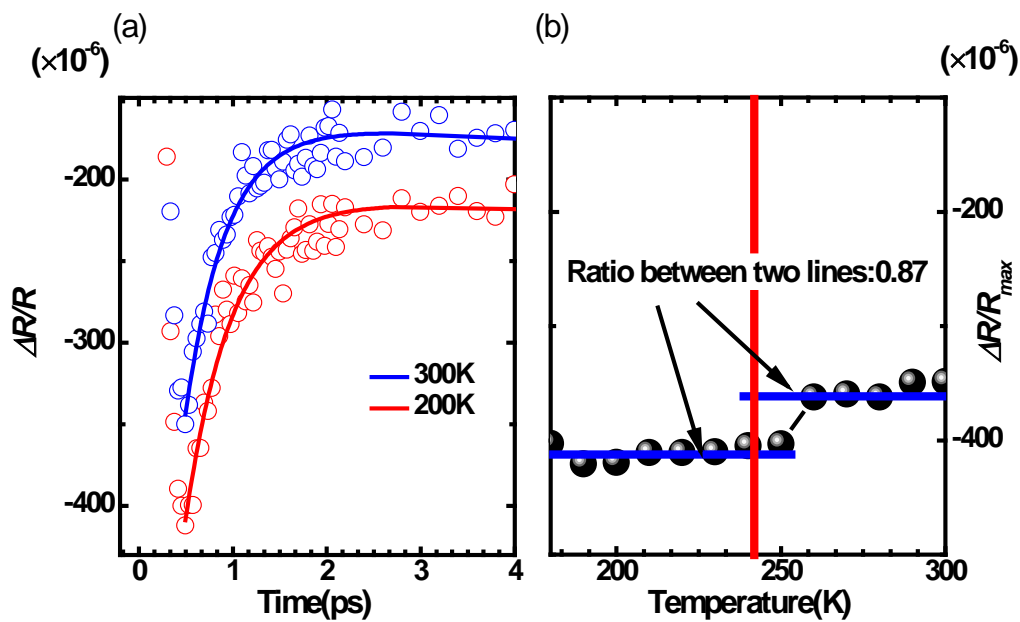

**Figure 4. Peak amplitude of the transient reflectivity change near $T_N$ after 1.5 eV excitation.** Temperature-dependent transient reflectivity change at 1.1 eV after photoexciting the $Fe^{2+} \rightarrow Fe^{3+}$ intralayer charge transfer excitation at 1.5 eV.

**Methods**

**Experimental setup and sample preparation.** Our femtosecond optical pump-probe spectroscopy system is based on a 75 femtosecond (fs), 250 kHz repetition rate Ti:sapphire chirped pulsed amplifier operating at 800 nm (~1.5 eV) and seeding an optical parametric amplifier (OPA) that allows us to tune the photon energy. A delay line allows us to vary the optical path difference between the pump (200 μm diameter) and probe (100 μm diameter) beams, which are then focused to the same spot on the sample. Temperature-dependent transient reflectivity changes were obtained in reflection with cross-polarized pump and probe beams in a >10:1 power ratio (the pump fluence is 76 μJ/cm$^2$, which photoexcites ~0.007 electrons/unit cell), incident at an angle of less than 10$^o$ to the hexagonal c axis of the crystal. The probe photon energy was 1.1 eV in all experiments (examining $E_\uparrow$), and the pump photon energies were 1.1 and 1.5 eV (photoexciting $E_\uparrow$ and $E_{t\rightarrow}$, respectively). It is worth mentioning that pump fluence-dependent measurements revealed that only the amplitude of the *ΔR/R* signal changes linearly with fluence, with no changes in the measured dynamics, for both 1.1 and 1.5eV excitation. Furthermore, at the fluence used here, the maximum temperature increase due to laser heating is <8 K, which should not significantly affect the measured dynamics, and the sample completely recovers in the 4 μs time interval between amplifier pulses. Finally, the LuFe$_2$O$_4$ single crystal used in this study was grown by the floating zone method as described in ref. [15], with its surface normal to the c-axis.

**Theoretical background.** Quantum charge fluctuations originate from hopping of an electron between two spatially separated potential minima. When an electron is localized in one potential well, the system becomes electrically polarized [20, 21]. If we consider two energetically non-degenerate states localized at two different lattice sites and include the double exchange interaction in the system, the Hamiltonian can be expressed as

$$H^0 = \varepsilon_1 C_1^+ C_1 + \varepsilon_2 C_2^+ C_2 - t_{12} C_1^+ C_2 - t_{21} C_2^+ C_1, \qquad (M1)$$

where $\varepsilon_i$ is the atomic energy, $C_i^+$ and $C_i$ are creation and annihilation operators at the $i$th site ($i=1, 2$), and $t_{ij}= t_{ji}$ is a effective hopping matrix element accounting for the hopping between sites. This matrix element is governed by the double exchange mechanism, which relates the probability of an electron hopping between two atoms to the angle $\theta_{ij}$ between the local core spins $\mathbf{S}_i$ and $\mathbf{S}_j$ [22, 23]. This mechanism has most frequently been used to explain the metallic properties of colossal magnetoresistive manganites [22], but also applies here to $LuFe_2O_4$, since the ferrimagnetic spin order existing below $T_N$ influences electron hopping (and therefore charge fluctuations of the charge-ordered state), which can in turn change the dielectric properties of the system. In other words, since the electronic charge is never completely localized in insulators, the degree of delocalization depends on the effective hopping amplitude given by the double exchange mechanism. Since magnetic ordering suppresses this hopping amplitude for the $\mathbf{E}_{b\rightarrow}$ and $\mathbf{E}_\downarrow$ transitions, we expect electrons in $LuFe_2O_4$ to be more localized, stabilizing charge order.

For small $t_{ij}$ values ($t_{ij}<<\Delta$, where $\varepsilon_2 - \varepsilon_1 \equiv \Delta$ and $t_{12}=t_{21}$) and $\varepsilon_2 > \varepsilon_1$, most of the charge will be localized at site 1, with a small fraction $\delta q = \left(\dfrac{t_{ij}}{\Delta}\right)^2$ of delocalized charge remaining at site 2. Because the electric polarization $\mathbf{P}$ is proportional to the difference of charge densities between sites, $\rho_2 - \rho_1$ (where $\rho_2$ and $\rho_1$ are the electron densities at sites 2 and 1, respectively), any change in the delocalized charge at site 2 causes a change in $\mathbf{P}$.

When an external electromagnetic field ($E_0 \cos \omega t$) is applied to the system, it will introduce a small perturbation $H^1 = exE_0 \cos \omega t$ into the Hamiltonian, inducing a site-to-site transition (where $e$ is the electron charge and $x$ is the distance between the two sites). We can use Fermi's golden rule [33] to calculate the probability of transitions between both sites (corresponding to quantum charge fluctuations), which is found to be proportional to the delocalized charge on site 2, $\delta q$, through $H_{12}^2 = |\langle \Psi_2 | H^1 | \Psi_1 \rangle|^2 \sim (\frac{t_{12}}{\Delta})^2 \sim \delta q$ ( where $|\Psi_1\rangle$ and $|\Psi_2\rangle$ are the ground and excited states of $H^0$, respectively). Note that the transition rate is proportional to the extent of charge delocalization in the ground state of the system. Since the photoinduced change in reflectivity at the absorption peak is proportional to changes in the absorption under the conditions of our experiment [31], which, in turn, is proportional to $H_{12}^2$ and $\delta q$, we can relate our transient reflectivity measurements to the amount of delocalized charge and thus the polarization **P**. This then establishes that we can use our ultrafast optical measurements to reliably photoexcite and probe quantum charge fluctuations in LuFe$_2$O$_4$.

We first calculate $\Delta E_\uparrow$, i.e. the pump-induced difference in the interlayer charge transfer energy based on the new Coulomb energy after exchanging Fe$^{2+}$ and Fe$^{3+}$ ions, using equation (1) and only considering four ions: two excited by the pump and two examined by the probe after excitation (Figure 2 in our manuscript). This was done by exchanging a Fe$^{2+}$ and a Fe$^{3+}$ ion either between the bottom and top layers (corresponding to absorption of a 1.1 eV pump photon( figure 2(a))) or between two sites in the top layer (corresponding to absorption of a 1.5 eV pump photon(figure 2(b)). This result was then used to calculate the change in absorption as described above, from which we calculated the variation of $\Delta R/R_{max}$ with temperature after both 1.1 eV and 1.5 eV photoexcitation for comparison to our experimental data. It is worth noting that if either of the ions that absorb a pump photon is involved in the subsequent absorption of a probe photon, we find that the resulting $\Delta E_\uparrow$ is much larger than the probe bandwidth (~13 meV) and thus does not contribute to the observed absorption change. Finally, including more than 47 $Fe^{2+}$ electrons in this calculation resulted in an insignificant reflectivity change, since there is almost no change in $E_\uparrow$ due to $Fe^{2+}$ ions far from the $Fe^{2+}$ and $Fe^{3+}$ sites that participate in the photoinduced transition.

,